\documentclass[11pt]{article}
\usepackage{moriond,epsfig}

\def\be{\begin{equation}}
\def\ee{\end{equation}}
\def\bea{\begin{eqnarray}}
\def\eea{\end{eqnarray}}

\begin{document}

\vspace*{4cm}
\title{
LENSED SUBMILLIMETRE-WAVE FOREGROUNDS AND THE CMBR}

\author{ A.W. BLAIN }

\address{Cavendish Laboratory, Madingley Road, Cambridge,\\ 
CB3 0HE, United Kingdom}

\maketitle\abstracts{
Samples of high-redshift galaxies are easy to select in the
millimetre/submillimetre (mm/submm) waveband using sensitive 
telescopes, because their flux density--redshift relations are 
expected to be flat, and so the selection function is almost 
redshift-independent at redshifts greater than 0.5. Source counts 
are expected to be very steep in the mm/submm waveband, and so 
the magnification bias due to gravitational lensing is expected to be 
very large, both for lensing by field galaxies and for lensing by 
clusters. Recent submm-wave observations of lensed images in 
clusters have constrained the submm-wave counts directly for the 
first time. In the next ten years our knowledge of galaxy evolution in 
this waveband will be greatly enhanced by the commissioning of 
sensitive new instruments and telescopes, including the CMBR 
imaging space mission {\em Planck Surveyor}. This paper highlights 
the important features of gravitational lensing in the submm waveband 
and discusses the excellent prospects for lens searches using 
these forthcoming facilities.
}


\section{Distant dusty galaxies}

The flux density--redshift relation for distant dusty star-forming and active 
galaxies in the submm waveband is expected to be approximately flat at 
redshifts in the range $0.5 \le z \le 10$, because of the large negative 
$K$--correction obtained when the steep modified thermal dust spectra of 
these galaxies, which peak in the rest-frame far-infrared waveband, are 
redshifted into the observed mm/submm waveband.\cite{BL} Consequently, the 
faint submm-wave source counts are expected to be very steep as compared 
with the faint counts in other wavebands. The surface density of such galaxies at 
mJy flux densities was uncertain by up to three orders of magnitude\,\cite{BL96} 
until six sources were detected in the fields of two rich clusters of 
galaxies\,\cite{Cluster} in 850-$\mu$m observations using the SCUBA bolometer 
array receiver at the James Clerk Maxwell Telescope (JCMT) last year;\,\cite{SIB} 
it is now known to within a factor of about two.\footnote{
This work has benefited greatly from the results of observations made using 
SCUBA at the JCMT in collaboration with Ian Smail, Rob Ivison and 
Jean-Paul Kneib.}
The brightest of these sources 
has now been identified with an obscured AGN/starburst galaxy at 
$z=2.803$.\cite{I+7} 
These observations allowed the source confusion noise that will affect 
mm/submm-wave CMBR observations, made by either balloon-borne 
experiments or the {\em Planck Surveyor} space mission,\cite{RB} to be 
estimated directly for the first time.\cite{BIS} Lensing is not expected 
to modify the predicted source confusion noise significantly.

\section{Gravitational lensing}

Because of the large negative $K$--correction for galaxies in the submm
waveband, the redshift distribution of faint galaxies is expected to extend 
out to much larger redshifts as compared with other wavebands. Hence the 
gravitational lensing cross section for a typical source is also expected to be 
large as compared with that in other wavebands. In addition, the steep 
submm-wave counts are expected to be associated with a large positive 
magnification bias\,\cite{GGlens} over a wide range of flux densities; that is 
lensing is expected to produce a considerable increase in the surface density 
of detectable sources. The effects of lensing by both
galaxies\,\cite{GGlens,Grenoble} and clusters,\cite{Cluster,SZ} and the 
consequences of different world models for the predictions,\cite{Cos,New} are 
described in detail elsewhere. In the light of the first observed 850-$\mu$m 
counts\,\cite{SIB} realistic estimates of the observability of lensed galaxies using 
existing and future instruments can be made.\cite{BIS,SZ,Planck} The details of 
the observations\,\cite{SIB} and lensing
calculations\,\cite{Cluster,GGlens,Grenoble,Cos} are not discussed in this short 
contribution. 

The properties of galaxy--galaxy lenses in the field and images lensed by 
clusters are considered in Sections 3 and 4 respectively. The catalogues 
of point sources that will be detected in mm/submm-wave CMBR surveys, and 
especially in the all-sky {\em Planck Surveyor} survey,\cite{RB} will allow large 
samples of galaxy--galaxy lenses to be compiled, as summarized in Section\,5.

\section{Galaxy--galaxy lensing}

The submm-wave source counts are expected to be steep at 
observable flux densities, and so the corresponding magnification biases are 
predicted to be large. Predictions of the counts of both lensed and 
unlensed galaxies at wavelengths of 850 and 500\,$\mu$m, based on direct 
submm-wave observations,\cite{SIB} and existing models of both the lensing 
optical depth\,\cite{GGlens,Cos} and the evolution of galaxies\,\cite{BL96} are 
shown in Fig.\,1. The counts are compared with the 3$\sigma$ sensitivities of
long-term surveys made using a range of existing and future mm/submm-wave 
telescopes; SCUBA at the 15-m JCMT, a large ground-based interferometer array 
such as the MMA,\cite{MIA} the 3.5-m space-borne telescope 
{\em FIRST},\cite{FIRST} and the space-borne all-sky CMBR imaging satellite 
{\em Planck Surveyor}.\cite{RB} If the curves representing the predicted counts 
lie above the thin solid line labelled with the name of a particular telescope 
then that telescope is expected to detect at least one source in the survey. A 
more detailed description of these instruments and their likely levels of source 
confusion is given elsewhere.\cite{BIS}

The ratio of the numbers of lensed to unlensed galaxies with flux densities 
greater than about 50\,mJy -- the 3$\sigma$ sensitivity of the {\em Planck 
Surveyor}\,\cite{RB} -- is predicted to be greater than 1\%, which illustrates the 
very large magnification bias. The surface density of such lenses is expected to 
be between about 10$^{-2}$ and 0.1\,deg$^{-2}$. Hence, a large sample of lensed 
star-forming galaxies\,\cite{Grenoble,Planck} and quasars\,\cite{I+7} could be 
found in a practical survey using the next generation of submm-wave 
instruments. The 50-m single-antenna US/Mexican LMT/GTM
telescope\,\cite{LMT} will also have a 
formidable performance in a lens search. In a practical 
{\em Planck-Surveyor}-based survey\,\cite{Planck} up to $10^3$ lensed sources 
could be detected. The inclusion of a population of galactic disks in the 
population of lensing galaxies\,\cite{MB} can increase the optical depth to 
high-magnification lensing by a significant factor,\cite{BMM} and so this 
remarkably large number of lens detections expected in the {\em Planck 
Surveyor} survey may actually be a conservative estimate of the true number. 

\section{Lensing by clusters}

The counts on which the results in this paper are based were determined by 
submm-wave observations of two rich lensing clusters.\cite{SIB} The large 
$K$--corrections for background galaxies ensure that galaxies within the 
clusters at $z \ll 1$ are unlikely to contaminate the results,\cite{Cluster} and a 
sample of six clusters has now been observed.\cite{New} The magnification 
bias factor in the cores of rich clusters is predicted to be several hundred in the 
submm waveband, as compared with less than about ten in the optical 
waveband,\cite{Cluster} and so these observations offer considerable promise 
both for investigating the history of star and galaxy formation,\cite{BSI}
unbiased by the effects of dust extinction, and the values of cosmological 
parameters.\cite{Cos} Lensing is expected to increase the mean flux density 
and angular separation of point sources in the fields of clusters, which is
expected to increase the severity of source confusion noise in the fields of 
clusters as compared with the field when observed on arcminute scales
in the mm/submm waveband.\cite{SZ} Hence, if a very accurate measurement of 
the Sunyaev--Zel'dovich effect in clusters\,\cite{Lange} is required then the flux 
density due to confusing point sources may need to be subtracted from these 
observations, using arcsecond-resolution maps of the cluster obtained using 
either a large single-antenna telescope\,\cite{LMT} or a large 
interferometer.\cite{MIA}

\begin{figure}
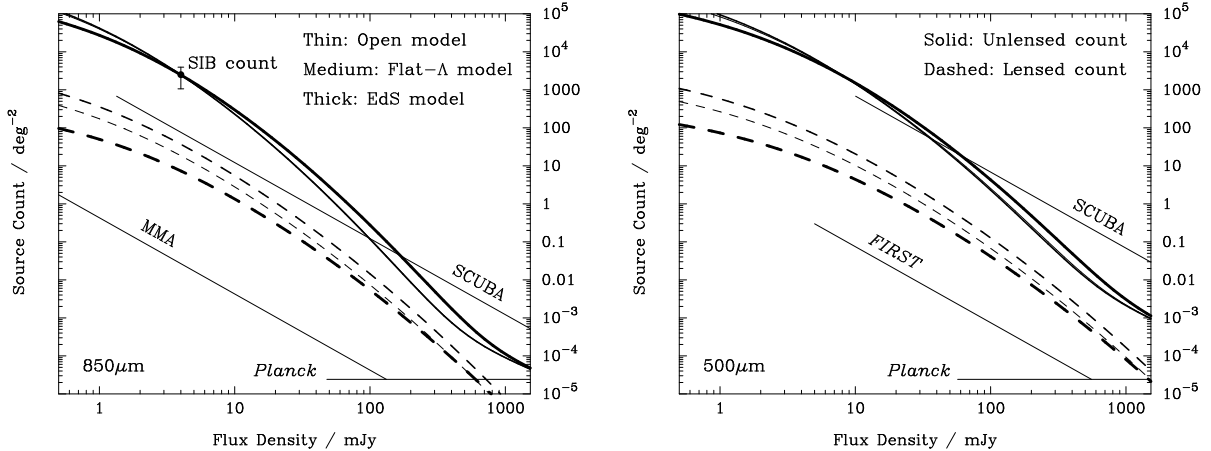

\psfig{figure=Moriond850.ps,height=3.0in,angle=-90}
\hskip5mm
\psfig{figure=Moriond500.ps,height=3.0in,angle=-90}
\caption{Predicted  
submm-wave source counts of both lensed (dashed lines) and 
unlensed (solid lines) galaxies at wavelengths of 850\,$\mu$m (left) and 
500\,$\mu$m (right) in three different world models. 
The Einstein--de Sitter (EdS) 
model is represented by thick curves; the Flat-$\Lambda$ model, with 
$\Omega_0=0.3$ and $\Omega_\Lambda=0.7$, is represented by curves of 
medium thickness, and the Open model, with $\Omega_0=0.3$ and 
$\Omega_\Lambda=0$, is represented by the thin curves. The counts of unlensed 
galaxies in the Open and Flat-$\Lambda$ models are almost indistinguishable; 
they lie below the counts in the EdS model. 
The predicted counts are consistent 
with counts and upper limits to counts derived from current 
observations made at 
wavelengths of 850\,$\mu$m (SIB$^{4}$), 2.8\,mm$^{19}$ and 
175\,$\mu$m$^{20}$ and with the observed intensity of background 
radiation.$^{11}$ In the standard notation$^{2}$ $p=3$, 
$z_0=7$; in the 
EdS, Flat-$\Lambda$ and Open models $z_{\rm max}$ takes the values 2.6, 2.87 
and 3.0 respectively. The sensitivities of existing and future instruments are 
included for comparison. If the counts lie above the lines representing 
instrumental sensitivities then 3$\sigma$ detections are expected in a survey. 
An 
integration time of 200\,h is assumed for a SCUBA survey; 1000\,h is assumed 
for {\em FIRST}$^{15}$ and the MMA.$^{14}$ The nominal 14-month 
{\em Planck Surveyor}$^{6}$ mission is assumed. Hubble's constant 
$H_0=50$\,km\,s$^{-1}$\,Mpc$^{-1}$.
} 
\end{figure}

\section{Lens surveys in the mm/submm waveband}

Although the SCUBA camera at the 15-m JCMT is much more sensitive than 
previous submm-wave instruments, it is not expected to detect a large number 
of galaxy--galaxy lenses, as shown in the left-hand panel of Fig.\,1 and 
discussed elsewhere.\cite{GGlens} The line represented the sensitivity 
limit of SCUBA clearly lies above the expected count of lensed sources.

However, the prospects for using future mm/submm-wave telescopes, 
{\em FIRST}, {\em Planck Surveyor}, the MMA and the LMT/GTM to detect and 
study lensed galaxies and quasars appear to be excellent. In Fig.\,1 the lines 
representing the sensitivity limits of {\em FIRST}, {\em Planck Surveyor} and the 
MMA all lie significantly below the expected counts of lensed sources. The
BOLOCAM receiver at the LMT/GTM will operate at wavelengths longer than
1.1\,mm, and although not shown in Fig.\,1, its performance is expected to be 
similar to that of the MMA. Mm/submm-wave surveys using these instruments 
will all produce a large catalogue of sources, with flux densities in the range 
10--100\,mJy for the space-borne surveys,\cite{Grenoble,Planck} in the 
range 1--10\,mJy for the LMT/GTM and potentially at 100-$\mu$Jy flux densities 
for the MMA. However, only the MMA has the potential to resolve the multiple 
image components of a lensed system directly. The sample of sources
detected using other telescopes will require high-resolution follow-up imaging 
using an interferometer array, and final spectroscopic confirmation using large 
near-infrared/optical telescopes in order to decide whether they are lensed or
not. These follow-up observations would require a concerted campaign of 
observing lasting for several months; however, they are practical.\cite{Planck}
A catalogue containing of order 1000 galaxy--galaxy lenses will be the final 
product of future mm/submm-wave lens surveys.  

\section{Conclusions}

\begin{enumerate} 
\item 
The surface density of distant galaxies in the submm waveband, and therefore 
the expected surface density of gravitational lenses and the effects of source 
confusion in future observations, is now known with reasonable accuracy.
\item 
The {\em Planck Surveyor} survey and surveys using other forthcoming 
mm/submm-wave telescopes will produce catalogues of distant sources that 
will be of great interest for studies of galaxy evolution. Lensed galaxies and 
quasars will be detected with an efficiency of up to 10\% in these surveys, and a 
sample of order 1000 lenses could be compiled.  
\end{enumerate}


\end{document}